\documentstyle[12pt]{article}

\def\be{\begin{equation}}
\def\ee{\end{equation}}
\def\a{\alpha}
\def\b{\beta}
\def\ra{\rangle}
\def\la{\langle}

\def\sin{\mbox{sin}}
\def\cos{\mbox{cos}}
\def\exp{\mbox{exp}} 
\def\sign{\mbox{sign}}

\def\w{\omega} 
\def\s{\sigma}
\def\l{\lambda}
\def\D{\Delta} 
\def\d{\partial}
\def\L{\Lambda}
\def\tp{t^{\prime}} 
\def\xip{\overline{\xi}}

\begin{document}
 
\begin{center}
{\bf Bosonization and  correlators of the one-dimensional Hubbard model.}
\end{center}
\vspace{0.2in}
\begin{center}

{\large A.A.Ovchinnikov}

\end{center}   

\begin{center}
{\it Institute for Nuclear Research, RAS, Moscow}
\end{center}   
 
\vspace{0.3in}

\begin{abstract}

We present simple derivation of the Luttinger liquid relation for the 1D Hubbard model 
both for finite $U$ and in the $U=\infty$ limit. We describe the simple solution of the 
Hubbard model in the infinite repulsion limit and use it to calculate the correlators of the model in this limit in a simple and a physical way using the Bosonization technique. 
We then calculate the asymptotics of the correlators of the model at arbitrary $U$ 
through the single parameter, which can be calculated from the Bethe Ansatz solution. 
Our derivation of the critical exponents is simple and allows one to express different 
physical operators of the Hubbard model through the charge and spin Bose fields in a 
direct and a physically transparent way.

\end{abstract}

\vspace{0.4in}

{\bf 1. Introduction}

\vspace{0.3in}

The one-dimensional (1D) Hubbard model is an excellent testing ground for various ideas in many fields of the condensed matter physics such as for example the high $T_c$ superconductivity. Besides it can be realized experimentally in many substances such as poliacetilen for example. The exact solution of the 1D Hubbard model by means of the Bethe Ansatz \cite{L} allows one to study various properties of the model such as the energy of the ground and the excited states, thermodynamics and the asymptotic behaviour of the correlation functions which was first presented in Ref.\cite{K}. These calculations based on the Conformal Field Theory where continued and simplified by different methods and for various special values of the parameters (on-site repulsion $U$, electron density $n$ and the magnetic field $h$) by several authors. For example, in Ref.\cite{AR} the simple derivation of the critical exponents in the limit $U=\infty$ was presented and in Ref.\cite{Penc} the connection of the low-energy properties of the 1D Hubbard model with the Luttinger liquid 
model was pointed out. 

Inspite of these achievements the derivation of the correlators (critical exponents) by a more simple and physically more transparent methods is of interest. In the present paper we use Bosonization and the Luttinger liquid concept to calculate the correlators in the model in a different simple and physically more transparent way. 
Let us explain the main goals of the present paper in detail. There are two different ways to write down the Luttinger Liquid relations for the Hubbard model which corresponds to the two different (but related) 
sets of the quantum numbers. The Bosonization was already used to study the correlation functions 
starting from the Luttinger Liquid relation for the quantum numbers, which are more natural in the weak 
coupling limit (for example, see \cite{S}, \cite{G} and references therein). 
However, in the strong coupling limit the set of the quantum numbers associated with the Bethe Ansatz 
solution of the model is more natural. Therefore the first goal of the present paper is to derive the Luttinger Liquid 
relation for these quantum numbers {\it without} reffering to the Bethe Ansatz solution. 
This can be done at arbitrary interaction strength up to a single unknown parameter. In the infinite coupling 
limit this parameter can be fixed without using the Bethe Ansatz solution. So in this limit we derive the 
Luttinger Liquid relation without reffering to the Bethe Ansatz  solution.
The second goal is to calculate the correlators of the Hubbard model via Bosonization from the 
Luttinger Liquid relation for these quantum numbers. 
One is able to find the expressions for the lattice operators through the charge and the spin Bose 
fields corresponding to these quantum numbers from the physical arguments. 
This problem was not solved before. 
We calculate the critical exponents for the model at zero magnetic field $h=0$ and an arbitrary $U$, $n$, but the special attention is paid to the case of the infinite repulsion $U=\infty$. Here we show in detail how the Bosonization and the Luttinger liquid low-energy theory lead to the known critical exponents in a very simple and beautiful way. More specifically the main problem of Ref.\cite{K} - the expression of the physical operators in terms of the Bose- fields is solved here naturally with the help of Bosonization. In the case $U=\infty$ our method allows to avoid the rather complicated calculations \cite{Parola} and the numerical calculations \cite{OS} due to the rather simple wave function of the model of the type presented in \cite{OS},\cite{O}. 
This is the main goal of the present paper.

In Section 2 we describe the intuitive solution of the Hubbard model at $U=\infty$ (infinite repulsion). The solution can be obtained in two different ways: the simple intuitive solution specific for $U=\infty$ and as a limit of the Bethe Ansatz wave function and the equations. 
In Section 3 we derive two types of the Haldane - Luttinger relations \cite{H} corresponding to the two different sets of the quantum numbers and valid both at $U=\infty$ and an arbitrary $U$. Using the Bosonization  approach we calculate  the asymptotics of the correlators in Section 4. 
This also can be done for two different sets of the quantum numbers. 
We establish the relation between them and show their equivalence. We 
present the results for the equal-time  correlators at an arbitrary $U$ and $n$.  
Finally in the Appendix we briefly discuss the calculation of the exponents from the 
Bethe Ansatz equations.

\vspace{0.6in}

{\bf 2. Infinite repulsion solution of the Hubbard model.}

\vspace{0.2in}

Let us describe the solution of the 1D Hubbard model with infinite repulsion. Initially the Hamiltonian of the model has the following form: 
\be
H=-t\sum_{\la ij \ra\s}\left(c_{i\s}^{+}c_{j\s} + c_{i\s}^{+}c_{j\s}\right)+
U\sum_{i}n_{1i}n_{2i}+ h\sum_{i}(n_{1i}-n_{2i}), 
\label{H}
\ee
where $c_{i\s}^{+} (c_{i\s})$ are the creation (annihilation) operators of electrons and 
$n_{\s i}=c_{i\s}^{+} c_{i\s}$ are their numbers at the site $i$. 
Here we put the parameter $t=1$, and the index $\s$ takes two values 
$\s=1,2=\uparrow, \downarrow$. The $h$ is the magnetic field which we take equal to zero $h=0$. The Hamiltonian (\ref{H}) depends on the two parameters: the repulsion $U$ and the total number of electrons $N_e=N_1+N_2$. We denote by $M$ the total number of the spin-up electrons, $M=N_1$ and by $L$ the length of the chain. The sum in eq.(\ref{H}) is over the pairs of the nearest neighbor sites, and the periodic boundary conditions are implied.  
To consider the limit $U\rightarrow\infty$, when the double occupied sites are forbidden,  
it is convenient to express the Hamiltonian (\ref{H}) 
in terms of the fermionic spinless creation and annihilation operators of the holes (empty sites)
$(c_i^{+}, c_i)$ defined starting from the ferromagnetic state 
$|F\ra=\prod_{i}c_{2i}^{+}|0\ra$.  The up-spin electrons are described by the hard-core Bose 
operators $(b_i^{+}, b_i)$. If we denote by $n_i$ the number of bosons ($n_i=b_i^{+}b_i$), the Hamiltonian (\ref{H}) in the sector without the double occupied sites takes the form: 
\be
H=-\sum_{\la ij \ra}c_i^{+}c_j\left((1-n_i)(1-n_j)+b_j^{+}b_i\right)+ h.c.
\label{Hbc}
\ee
Let us seek the wave function of the Hamiltonian (\ref{Hbc}) in the following form: 
\be
\psi(i_1,\ldots i_N|l_1,\ldots l_M)=\psi_{0}(i_1,\ldots i_N)\phi(\l_1,\ldots \l_M), 
\label{psi}
\ee
where $i_{\a}$ are the coordinates of $c$- particles ($N=N_h$ is the number of holes), $l_{\a}$ 
are the coordinates of the up-spin particles ($b$- particles) and 
$\l_{\b}$ are the coordinates of the spin bosons on a "superlattice" which consist of 
$L_1=L-N$ lattice sites which are not occupied by the holes ($L_1=N_e$ is the number of electrons), 
\[
\l_{\a}=l_{\a}-\sum_{\b=1}^N\theta(l_{\a}-i_{\b}), ~~~~~\a=1,\ldots M. 
\]
Clearly, if $\psi_0$ is the eigenstate of the Hamiltonian (\ref{Hbc}) in the sector with the maximal total spin, the wave function (\ref{psi}) is the eigenfunction of the Hamiltonian (\ref{Hbc}) for an arbitrary function $\phi(\l_1,\ldots \l_M)$. 
Now let us take into account the periodic boundary conditions for the holes. Imagine the hole jumps from the site $1$ to the site $L$. Then we get the shift of the coordinates of spins on a superlattice $\l_{\a}\rightarrow\l_{\a}+1$ and we obtain the additional phase shift $q$: 
\be
\psi_{0}(i_1+L, \ldots i_N)=\exp(iq)\psi_{0}(i_1,\ldots i_N), 
\label{kq}
\ee
where $q$ - is the total momentum corresponding to the spin wave function $\phi(\l_1,\ldots \l_M)$: 
\be
\phi(\l_1+1,\ldots \l_M+1)=e^{iq}\phi(\l_1,\ldots \l_M). 
\label{q}
\ee
For the Hamiltonian (\ref{H}) one can perform the perturbation theory in $1/U$ at large 
$U\rightarrow\infty$. From this fact it follows that for the Hubbard model the degenacy is removed  
and the function $\phi(\l_1,\ldots \l_M)$ is the eigenstate of the Antiferromagnetic Heisenberg model 
(XXX- spin chain) on the superlattice $(1,\ldots L_1)$ \cite{OS}.  
From the Hamiltonian (\ref{Hbc}) and the equations (\ref{kq}), (\ref{q}) one can see that 
the function $\psi_{0}(i_1,\ldots i_N)$ is the Slater determinant corresponding to the momenta 
$k_1,\ldots k_N$, which are equal to 
\be 
k_{\a}L=2\pi n_{\a}+q, ~~~~q=\sum_{\b}q_{\b}, ~~~~q_{\b}=\frac{2\pi m_{\b}}{L_1}. 
\label{K}
\ee
The energy equals $E=-2\sum_{\a}\cos(k_{\a})$ and the equations (\ref{K}) give the 
complete solution of the problem. Here the quantum numbers $n_{\a}$ are integer and 
$m_{\a}$ - are integer or half-integer depending on the number of bosons $M$ according
to the well known solution of the Heisenberg model on a superlattice. The more detailed 
treatment can be found in Ref.\cite{O}. 
The correspondence of our solution (\ref{K}) with the large $U$ - limit of the Bethe Ansatz 
solution (for example, see \cite{OS}) can be established via the following redefinition 
of the quantum numbers. Namely one can perform the following shift of $n_{\a}$ and $m_{\b}$: 
\be
n_{\a}\rightarrow n_{\a}-M/2, ~~~~~m_{\b}\rightarrow m_{\b}+L_{1}/2. 
\label{mn}
\ee
The new quantum numbers coincide with the usual Bethe Ansatz quantum numbers located 
symmetrically around zero. We get for the new quantum numbers: $n_{\a}$- is integer 
(half-odd integer) for $M$- even (odd) and $m_{\b}$- is integer (half-odd integer) for 
$L_1-M$- odd (even), which corresponds exactly to the Bethe Ansatz solution.

\vspace{0.4in}

{\bf 3. Derivation of the Luttinger Liquid relation.}

\vspace{0.2in}

The minimal energy of the eigenstates in sectors with a small deviation of the particle  numbers which is of order $\sim 1/L$ is often called as a Luttinger liquid relation \cite{H}. 
Here we derive these relations for the model (\ref{H}) using two different sets of the 
quantum numbers. At $U=\infty$ one is able to fix the parameters of these relations 
$\xi_c$, $\xi_s$, while at an arbitrary $U$ they depend on a single parameter $\xi$, 
which should be determined from the Bethe Ansatz equations.  
The eigenstates are characterized either by the numbers $\D N_{1,2}^{(1)}$, 
$\D N_{1,2}^{(2)}$ which is the numbers of the additional right or left moving particles 
of the spin 1 or 2, or by the quantum numbers $\D N_{c,s}$, $\D Q_{c,s}$ associated 
with the sets $\{n_{\a}\}$, $\{m_{\b}\}$ in the equations (\ref{K}). From 
$\D N_{1,2}^{(\s)}$, $\s=1,2$ one can define the charge and spin quantum numbers 
$\overline{\D N}_{c,s}$, $\overline{\D Q}_{c,s}$ in the usual way 
which are related to $\D N_{c,s}$, $\D Q_{c,s}$. In fact the usual spin and charge Bose 
fields correspond to the linear combination of the form 
\be
\overline{\D N}_{c,s}=(1/\sqrt{2})(\D N_1\pm\D N_2),  ~~~~
\overline{\D Q}_{c,s}=(1/\sqrt{2})(\D Q_1\pm\D Q_2),  
\label{def}
\ee  
where the quantum numbers $\D N_{\s}$ and $\D Q_{\s}$ for two different species of 
fermions $\s=1,2$ are equal to 
\be
\D N_{\s}=\D N_1^{\s}+\D N_2^{\s}, ~~~~\D Q_{\s}=\D N_1^{\s}-\D N_2^{\s},~~~~
\s=1,2. 
\label{defp}
\ee
Let us begin with the first set of the quantum numbers. In this case the energy has the form: 
\be
 \D E=\frac{\pi v_{c}}{2L}\left(\xi(\overline{\D N}_c)^2+
\frac{1}{\xi}(\overline{\D Q}_c)^2\right) + 
\frac{\pi v_{s}}{2L}\left((\overline{\D N}_s)^2+(\overline{\D Q}_s)^2\right).
\label{12}
\ee
Here the Luttinger Liquid parameters  $\xip_c$ and $\xip_s$ corresponding to the first set of 
the quantun numbers and defined in the standard way are equal to 
$\xip_c=\xi$, $\xip_s=1$,  
where the parameter $\xi=2$ for $U=\infty$ ($(\xip_c,\xip_s)=(2,1)$) and $\xi=1$ for 
$U=0$. The value $\xip_s=1$ follows from the $SU(2)$ invariance of the model (\ref{H}). 
For the quantum numbers (\ref{K}) the relation has the similar form: 
\be
\D E=\frac{\pi v_{c}}{2L}\left(\xi_{c}(\D N_c)^2+
\frac{1}{\xi_{c}}(\D Q_{c}+\D Q_{s}/2)^2\right) + 
\frac{\pi v_{s}}{2L}\left(\xi_{s}(\D N_{s}-\D N_{c}/2)^2+
\frac{1}{\xi_{s}}(\D Q_s)^2\right). 
\label{bc}
\ee
In general the parameters $\xi_c$, $\xi_s$ are equal to 
$\xi_c=\xi/2$, $\xi_s=2$.  Thus at $U=\infty$ we obtain $(\xi_c,\xi_s)=(1,2)$. 
To be precise let us repeat once more the main results of this Section for the equations 
(\ref{12}) and (\ref{bc}): 
\[
(\xi_c,\xi_s):~~~ (\xi,1)\rightarrow(\xi/2,2), ~~~~(2,1)\rightarrow(1,2) ~~~
(U=\infty), 
\]
where the arrows mean the transition from eq.(\ref{12}) to eq.(\ref{bc}). 
At finite $U$ the parameter $\xi$ varies from $\xi=1$ at $U=0$ to $\xi=2$ at 
$U=\infty$ ($1\rightarrow 2$).
It is one of the goals of the present
paper to derive the relation (\ref{bc}) from the solution presented in Section 2. 

The equivalence of the expressions (\ref{12}) and (\ref{bc}) can be shown in the following 
way. It is sufficient to consider the case $U=\infty$. First, with the help of the operators 
introduced in Section 2 we can calculate the total number of electrons with the spin 
1 and 2, i.e. $\D N_{1}^{(1)}+\D N_{2}^{(1)}$ and $\D N_{1}^{(2)}+\D N_{2}^{(2)}$. 
Second, we calculate the momenta (the currents) carried by the electrons of the type 
1 and 2: $\D N_{1}^{(1)}-\D N_{2}^{(1)}$, $\D N_{1}^{(2)}-\D N_{2}^{(2)}$. 
Here the following relation of the operators $b,c$ and $c_1,c_2$ valid at $U=\infty$ 
is useful: 
\be 
c_{i1}^{+}=b_i^{+}c_i, ~~~ c_{i1}=b_ic_i^{+}, ~~~c_{i2}^{+}=c_i, ~~~ 
c_{i2}=c_i^{+}. 
\label{equiv}
\ee 
Thus we obtain the following relations between the different quantum numbers \cite{Penc}: 
\[
\D N_{1}^{(1)}=\frac{1}{2}(\D N_s+\D Q_c+\D Q_s), ~~~ 
\D N_{2}^{(1)}=\frac{1}{2}(\D N_s-\D Q_c-\D Q_s), 
\]
\be
\D N_{1}^{(2)}=\frac{1}{2}(\D N_c-\D N_s+\D Q_c),  ~~~
\D N_{2}^{(2)}=\frac{1}{2}(\D N_c-\D N_s-\D Q_c).  
\label{1}
\ee 
With the help of the equations (\ref{1}) the equivalence of the relations (\ref{12}) and 
(\ref{bc}) can be easily proved.  

Now let us prove the relation (\ref{bc}) without reffering to the equations  (\ref{12}) and (\ref{1}) 
which is one of the goals of the present paper. 
It is necessary to consider the case $U=\infty$. 
The first (charge) term in the equation (\ref{bc}) at $\xi_c=1$ is nothing else but the energy of free fermions. The modification of the term $\sim(\D Q_c)^2$ comes from the additional 
phase shift (twist angle) 
$q\simeq(\pi/2)(\D N_1^{s}-\D N_2^{s}) =(\pi/2)\D Q_s$  
in the equation (\ref{K}). The same twist angle will appear in eq.(\ref{bc}) at an 
arbitrary $\xi_c$ ($U$). 
Here we have used the well known form of the Luttinger Liquid relation in a system with 
the twist angle. 
The modification of the second (spin) term in eq.(\ref{bc}) is connected with the fact 
that the Antiferromagnetic Heisenberg model is in fact defined on a superlattice. 
Initially we should write the Luttinger relation on a superlattice. 
Let us denote by $N_b(x)$ the number of the bosonic particles $b$ left to the site $x$. 
Clearly this number coincides with the same number on the superlattice which we denote 
by $N_b(x)|_{SL}$. Then we have: 
\[ 
N_b(x)|_{SL}=\frac{1}{2}x|_{SL}+\frac{1}{\sqrt{\pi}}\chi(x)|_{SL}, 
\] 
where $\chi(x)|_{SL}$- is the bosonic spin field on a superlattice. 
Since $x|_{SL}=x-N_h(x)$ the last equation can be rewritten in the form 
\be
N_b(x)|_{SL}=\frac{1}{2}x-\frac{1}{2}N_{h}(x) 
+\frac{1}{\sqrt{\pi}}\chi(x)|_{SL}=
 p_{F}x+\frac{1}{\sqrt{\pi}}(\frac{1}{2}\phi(x)+\chi(x)|_{SL}), 
\label{sl}
\ee
where $N_h(x)$ - is the number of holes left to the point $x$, $p_F$ - 
is the Fermi momentum 
and $\phi(x)$- is the bosonic field corresponding to the charge and connected with the 
number of holes. At the same time we can write for $N_b(x)$: 
\be 
N_b(x)=p_{F}x+\frac{1}{\sqrt{\pi}}\chi(x), 
\label{l}
\ee
where the field $\chi(x)$ is defined on the original lattice. Thus, comparing the equations 
(\ref{sl}) and (\ref{l}) we get the relation $\chi(x)|_{SL}=\chi(x)-\phi(x)/2$ which is 
equivalent to the substitution 
\[
\D N_s\rightarrow \D N_s-\D N_c/2 
\]
since the charge and the spin Bose fields $\phi(x)$ and $\chi(x)$ are correspond exactly 
to the quantum numbers $\D N_c$ and $\D N_s$. 
The value $\xi_s=2$ for $U=\infty$ is obvious since in this limit the second one of the 
Bethe Ansatz equations coincides with that of the Heisenberg Antiferromagnet. 
Thus the relation (\ref{bc}) is proved for the case $U=\infty$, $\xi_c=1$. 
Since the relations (\ref{1}) are valid at an arbitrary $U$ and $\xi_c$, $\xi_s$ are known 
from the equation (\ref{12}) through the single parameter $\xi$, the relation 
(\ref{bc}) is actually valid for an arbitrary $U$. 
At an arbitrary $U>0$ we know the form of the relation (\ref{bc}) from the equations 
(\ref{12}) , (\ref{1}), while for $U=\infty$ the value $\xi_{c}=1$ is derived from 
the wave function (\ref{psi}).   
Thus we have derived (\ref{bc}) both from the relation (\ref{12}) using the formulas (\ref{1}) 
and from the exact solution of the problem at $U=\infty$. The results are coincide. 
The derivation of the form of the relation (\ref{bc}) 
from the wave function (\ref{psi}) at $U=\infty$ is one of the main results of the present paper.

\vspace{0.6in}

{\bf 4. Asymptotics of the correlators at $U=\infty$.}

\vspace{0.2in}

Let us calculate the asymptotics of the correlators using the Bosonization approach (for example, see \cite{LP}). We transform the relations (\ref{12}), (\ref{bc}) into the  Hamiltonians corresponding to the two fields - charge field $\phi(x)$ and the spin field 
$\chi(x)$ according to the following rules: 
\[
\D N_c\rightarrow\d_1\phi(x),~~~\D Q_c\rightarrow\pi_{\phi}(x),~~~
 \D N_s\rightarrow\d_1\chi(x),~~~\D Q_s\rightarrow\pi_{\chi}(x), ~~\d_1=\d/\d x, 
\]
where $\pi_{\phi}$, $\pi_{\chi}$ - are the conjugated momenta, and express the physical 
operators of the model through these fields. 
First, for completeness let us present here the calculations \cite{S},\cite{G} related to the approach 
based on the relation (\ref{12}). In this case the 
expressions of the operators $c_1^{+}$, $c_2^{+}$ in terms of the (corresponding) 
fields $\phi(x)$, $\chi(x)$ are well known. Namely, one can write 
$c_{x\s}^{+}=e^{ip_{F}x}a_{\s}^{+}(x)+e^{-ip_{F}x}c_{\s}^{+}(x)$ and 
\[
(a_{1,2}^{+},c_{1,2})(x)=
e^{-i\sqrt{2\pi}((1/\sqrt{2})(\phi_{1,2}(x)\pm\int\pi_{1,2}(x)))}, 
\]
where the fields 
\be
\phi(x)=(1/\sqrt{2})(\phi_1(x)+\phi_2(x)),  ~~~~~ 
\chi(x)=(1/\sqrt{2})(\phi_1(x)-\phi_2(x)), 
\label{cs}
\ee
and we denote by 
$\int\pi_{1,2}(x)=\int^{x}dy\pi_{1,2}(y)$ the fields dual to $\phi_{1,2}(x)$. 
The calculations of different correlators are quite standard. Let us present the results 
for the critical exponents for an arbitrary parameter $\xi$. 
Since the first non-trivial term for density-density correlation coincides with the 
spin-spin correlator, we present here the results for the spin-spin correlator and the 
one-particle correlator connected with the momentum distribution. 
The results for the correlators $\la S^{+}_{x}S^{-}_0\ra$ and  
$\la S^{z}_{x}S^{z}_0\ra$ are coincide due to $SU(2)$ symmetry of the Hamiltonian 
(\ref{H}) and have the form: 
\be 
\la S^{+}_{x}S^{-}_0\ra\sim \cos(2p_{F}x)\frac{1}{x^{\a_{+}}}, ~~~
 \la S^{z}_{x}S^{z}_0\ra \sim\cos(2p_{F}x)\frac{1}{x^{\a_{z}}},  
\label{SS12}
\ee
where the critical exponents $\a_{+}$, $\a_{z}$ equal 
\be
\a_{+}=1/\xip_c+\xip_s=1/\xi+1=3/2~ (U=\infty), ~~
\a_{z}=1/\xip_c+1/\xip_s=1/\xi+1=3/2~ (U=\infty), 
\label{alphaSS}
\ee 
 where we have used the results of Section 3. The electron correlator has the form: 
\be
\la c_{1x}^{+}c_{10}\ra=\la c_{2x}^{+}c_{20}\ra\sim \cos(p_{F}x)/x^{\a_c}, 
\label{ee12}
\ee
where 
\be
\a_c=\frac{1}{4}(1/\xip_c+1/\xip_s+\xip_c+\xip_s)=\frac{1}{4}(1/\xi+\xi+2)=9/8 
~~(U=\infty). 
\label{alphaee}
\ee
Therefore the momentum distribution function has the following form: 
\be 
n_k=n_{p_F}-C|k-p_F|^{1/8}\sign(k-p_F) 
\label{nk}
\ee
at the infinite repulsion $U=\infty$.  
Thus the calculation of the correlators from the relation (\ref{12}) is completed. 

Now let us proceed with the calculation of the correlators at infinite $U$ using the 
wave function (\ref{psi}) which is the main goal of the present paper. 
Let us begin with the spin-spin correlator. For example consider 
the correlator $\la S^{+}_{x}S^{-}_0\ra$. The correlator for the Heisenberg model 
on a superlattice is well known. The leading term in the expansion of the operator 
$S^{+}_{x}$ in terms of the Bose fields is well known and has the form: 
\be
S^{+}_{x}=e^{i\pi x|_{SL}}e^{-i\sqrt{\pi}\int\pi_{\chi}(x)}. 
\label{Ssl}
\ee
Using the relation $x|_{SL}=x-N_{h}(x)$ we obtain the expression of the operator 
defined on the usual lattice: 
\be 
S^{+}_{x}=e^{i2p_{F}x}e^{i\sqrt{\pi}(\phi(x)-\int\pi_{\chi}(x))}. 
\label{Sl}
\ee
Using the relation (\ref{bc}) one obtains the correlator (\ref{SS12}) with the value of the 
exponent $\a_{+}=(1/\xi_c+\xi_s)/2=1/\xi+1=1/2+1=3/2$ in agreement with 
(\ref{alphaSS}) where we have used the values of $\xi_c$, $\xi_s$ from the equation 
(\ref{bc}). This method is very simple and allows one to calculate in a similar way the 
correlator of $z$- components of spin (\ref{SS12}). 

Now let us calculate the electron correlator (\ref{ee12}). To do it one should 
express the electron operators through the operators $b$ and $c$ (\ref{equiv}). 
Let us take the correlator $\la c_{2x}^{+}c_{20}\ra=\la c_{x}^{+}c_{0}\ra$ for example. 
One should express the hole operator $c_{x}^{+}$ through the bosonic fields. 
Here one should take into account the structure of the wave function (\ref{psi}). 
Apart from the usual terms with the fields shifted according to the equation (\ref{bc}), 
one should take into account the effect of the superlattice. In fact we notice that the 
operator $c_{x}^{+}c_{0}$ shifts the coordinates of spins on a superlattice $\l_{\a}$  
as  $\l_{\a}\rightarrow\l_{\a}+1$ for every $\l_{\a}$ located at the interval $(0,x)$. 
Thus we can write: 
\be
c^{+}_{x}=e^{i2p_{F}x}e^{i\sqrt{2\pi}((1/\sqrt{2})(\phi(x)+\int\pi_{\phi}(x)))}|_{shift}T(x), 
\label{c}
\ee
where the operator $T(x)$ shifts the coordinates $\l_{\a}$ at the interval $(0,x)$. 
One can represent this operator in the form $T(x)=\prod_{i=1}^{x}P_{i,i+1}$, where 
we denote by $P_{ij}$ the permutation operator acting at the sites $i$ and $j$. 
We note that the product of the two bosonic operators equals 
$b_{i}^{+}b_{i+1}\simeq(-1)b_i^{+}b_i=(-1)n_i$ which is valid up to the derivatives 
of the fields, which can be omitted. Then the permutation operator takes the form 
$P_{i,i+1}\simeq(1-2n_i)=(-1)^{n_i}$ and the operator $T(x)$ can be rewritten in the following form: $T(x)\simeq(-1)^{N_b(x)}=e^{-i\pi N_b(x)}$ where $N_{b}(x)$ - is 
the number of bosons at the interval $(0,x)$ on a superlattice. 
Now, according to Bosonization rules, one can represent this operator in the form: 
\be
T(x)\simeq e^{-ip_{F}x}e^{-i\sqrt{2\pi}\left((1/2\sqrt{2})\phi(x)+(1/\sqrt{2})\chi(x)
\right)}, 
\label{T}
\ee
where the second term in the exponent comes from the distance 
$x|_{SL}=x-N_{h}(x)$ and the term $\sim\chi(x)$ should be shifted according to  
eq.(\ref{bc}). Substituting the equation (\ref{T}) into eq.(\ref{c}) and taking into 
account the shift of the field $\int\pi_{\phi}(x)$, we obtain the operator of the hole 
\be 
c_{x}^{+}=e^{ip_{F}x}e^{i\sqrt{2\pi}\left((1/2\sqrt{2})\phi(x)+
(1/\sqrt{2})\int\pi_{\phi}(x)- (1/\sqrt{2})\chi(x)-(1/2\sqrt{2})\int\pi_{\chi}(x)\right)}. 
\label{final}
\ee
From eq.(\ref{final}) we immediately find the value of the critical exponent 
\be
\a_{c}=1/8\xi_c+\xi_c/2+1/2\xi_s+\xi_s/8= 
1/8+1/2+1/4+1/4=1+1/8  
\label{alphabc}
\ee 
in agreement with the equation (\ref{alphaee}). 
One can perform the similar calculations also for the correlator 
$\la c_{1x}^{+}c_{10}\ra$. Note also that we could perform the same calculations 
starting from the Bethe Ansatz basis after the shift (\ref{mn}). In this case the operator 
$e^{i\pi N_b(x)}$ appears from the very beginning and the operator $T(x)$ reduces 
to unity. Thus the correlators in the strong coupling limit are calculated which is the main 
result of the present paper.

\vspace{0.4in}

{\bf 5. Conclusion. }

\vspace{0.2in}

In conclusion, we derived the Luttinger liquid relation for the 1D Hubbard model 
at an arbitrary parameters $U$ and $n<1$ which apart from the velocities $v_c$, $v_s$ 
depends on a single parameter $\xi$, which can be calculated from the Bethe Ansatz 
equations. In particular at $U=\infty$ one can rigorously fix all the parameters entering 
the Luttinger liquid relation without reffering to the Bethe Ansatz solution.  
We have calculated the correlation functions of the Hubbard 
model using the Bosonization approach both in the electronic basis and in the strong 
coupling limit in the Bethe Ansatz basis starting from the simple factorized wave function 
(\ref{psi}) in this limit. 
Our derivation of the critical exponents is simple and allows one to express different 
physical operators of the Hubbard model through the charge and spin Bose fields in a 
direct and a physically transparent way. 
Let us stress once more that the goal of the present paper is not the derivation 
of the parameters of the Luttinger Liquid relation and the correlators from the 
Bethe Ansatz solution, but the simple and beautifull derivation of these quantaties 
without the Bethe Ansatz with the help of the Bosonization technique in the basis 
which is naturally connected with the Bethe Ansatz solution and especially usefull 
in the strong coupling limit.

\vspace{0.4in}

{\bf Appendix. }

\vspace{0.2in}

Let us present here the equation for the parameter $\xi_c=\xi/2$ in eq.(\ref{bc}) as a 
function of the parameters $U$, $n$ and derive its values in the limiting cases 
$U\rightarrow 0,\infty$ \cite{K} starting from the Bethe Ansatz solution. 
Let us denote by $k$ the momentum corresponding to the charge degrees of freedom 
(\ref{K}) and by $R(k)$ the corresponding density of roots. We also introduce the variable 
$t=\sin(k)$ such that $R(k)=\cos(k)R(t)$. At zero magnetic field excluding the BA 
equation for the spin degrees of freedom we get the equation 
\be
2\pi R(t)-\int_{-\L}^{\L}d\tp K(t-\tp)R(\tp)=S(t), ~~~~S(t)=1/\cos(k), 
\label{R}
\ee 
where $\L=\sin(k_0)$ is the cutoff. The function $K(t)$ in eq.(\ref{R}) is given by the 
following Fourier integral: 
\be 
K(t-\tp)=\int d\w e^{i\w(t-\tp)}\frac{1}{(1+e^{2u\w})},      
\label{Fourier}
\ee  
where $u=U/4$. We also introduce the dressed charge function $Z(t)$ difined by the 
equation 
\be
2\pi Z(t)-\int_{-\L}^{\L}d\tp K(t-\tp)Z(\tp)=2\pi. 
\label{Z}
\ee 
The simplest way to calculate the parameter $\xi$ is to consider the variation 
$\d n/\d\L$ which can be evaluated in two different ways. First 
from the equation (\ref{R}) we get 
\be
\left(2\pi-K\right)_{\L}\frac{\d R(t)}{\d\L}=\left(K(t-\L)+K(t+\L)\right)R(\L), 
\label{0}
\ee
where we denote by $\left(2\pi-K\right)_{\L}$ the corresponding integral operator 
at the interval $t\in(-\L,\L)$. The equation (\ref{0}) can be rewritten in the form: 
\be
\frac{\d R(t)}{\d\L}=\left(F(t-\L)+F(t+\L)\right)R(\L), ~~~
F(t-\tp)=\frac{K}{(2\pi-K)_{\L}}(t-\tp), 
\label{11}
\ee
where we formally introduced the inverse operator to the operator $(2\pi-K)_{\L}$. 
Now from the equation $n=\int_{\L}^{\L}dtR(t)$ we get for the derivative: 
\be
\frac{\d n}{\d\L}=R(\L)+R(-\L)+\int_{-\L}^{\L}dt\frac{\d R(t)}{\d\L}. 
\label{2}
\ee
Then using the definition (\ref{Z}) after some algebra 
we finally obtain the following simple relation: 
\be
 \frac{\d n}{\d\L}=2R(\L)Z(\L). 
\label{3}
\ee
Now one can show that the derivative (\ref{3}) can be found in a different way through 
the parameter $\xi$. In the framework of Bosonization the additional density $n$ is given by 
$\D\phi=\phi(L)-\phi(0)$, where $\phi(x)$ is the Bose field of the Luttinger model, so that 
$\d n/\d\L=\d\D\phi/\d\L$. At the same time since $2R(\L)d\L$- is the additional density 
of quasiparticles for the non-interacting field $\hat{\phi}(x)=\sqrt{\xi_c}\phi(x)$ we have 
$(1/2R(\L))\d\D\hat{\phi}/\d\L=1$. Thus we obtain the following equation for the 
parameter $\xi$: 
\be
\frac{1}{2R(\L)}\frac{\d n}{\d\L}=\frac{1}{\sqrt{\xi_c}}=Z(\L). 
\label{xi}
\ee
Then to obtain $\xi$ it is sufficient to solve the equation (\ref{Z}) for $Z(t)$ and find the 
value of $Z(\L)$. It is easy to show that at $U\rightarrow\infty$ $Z(\L)\rightarrow1$. 
At $U\rightarrow0$ one can use the Wiener-Hopf method (for example, see \cite{Y}) 
to obtain $Z(\L)=\sqrt{2}$. Thus at $U=0$ we obtain the value $\xi_c=1/2$, $\xi=1$ 
in agreement with the value obtained in the present paper.

\end{document}